\documentclass[11pt]{article}

\usepackage{hyperref}
\usepackage{authblk}
\pdfoutput=1

\begin{document}
\title{Bubble Rings Entrapment}
\author[1]{Marie-Jean Thoraval}
\author[1]{Sigurdur T. Thoroddsen\thanks{sigurdur.thoroddsen@kaust.edu.sa}}
\author[2]{Kohsei Takehara}
\author[2]{Takeharu Goji Etoh}
\affil[1]{Division of Physical Sciences and Engineering \& Clean Combustion Research Center,
King Abdullah University of Science and Technology (KAUST),
Thuwal, 23955-6900, Saudi Arabia.}
\affil[2]{Department of Civil and Environmental Engineering,
Kinki University, Higashi-Osaka, Japan.}

\maketitle

\begin{abstract}
We show how micro-bubble rings are entrapped under a drop impacting onto a pool surface.
This fluid dynamics video is submitted to the APS DFD Gallery of Fluid Motion 2012, part of the
65th Annual Meeting of the American Physical Society's Division of Fluid Dynamics (18-20 November, San Diego, CA, USA).
\end{abstract}

The video shows both results from numerical simulations using the {\it Gerris} open source code
and from ultra-high-speed video imaging using up to 1 million fps.

\begin{enumerate}
\item The numerical simulations show axisymmetric impacts where the drop is colored red and the pool is colored blue,
while the air is green.
\item In the experiments the impact is viewed through glass bottom through the shallow pool.  
The running numbers on the lower right on many of the videos clips are in micro-seconds.
\item The work shown in the video builds on our earlier numerical study and is currently being revised for publication
\begin{itemize}
\item{Thoraval, M.-J., Takehara, K., Etoh, T. G., Popinet, S., Ray, P.
Josserand, C., Zaleski, S. \& Thoroddsen, S. T. (2012),
Von K{\'a}rm{\'a}n vortex street within an impacting drop, {\it Phys. Rev. Lett.}, {\bf 108}, 264506. }
\item{ Thoraval, M.-J., Takehara, K., Etoh, T. G. and Thoroddsen,  S. T.,
Drop impact entrapment of bubble rings ({\it in preparation}).}
\end{itemize} 
\end{enumerate}
\end{document}